\documentclass[%
superscriptaddress,
pre, 
floatfix,
amsmath,
amssymb,
aps,
notitlepage
]{revtex4-1}

\usepackage[english]{babel}

\usepackage[hidelinks=true]{hyperref}
\usepackage{xcolor}
\definecolor{greenish}{rgb}{0.13,0.58,0.16}
\definecolor{reddish}{RGB}{174,12,48}
\definecolor{blueish}{rgb}{0.12, 0.56, 1.0}
\definecolor{magenta}{rgb}{0.8, 0.0, 0.8}

\hypersetup{
colorlinks   = true, 
urlcolor     = greenish, 
linkcolor    = magenta, 
citecolor   = blueish 
}
\usepackage{setspace}
\usepackage{caption}
\usepackage{siunitx}
\usepackage{subcaption}
\usepackage{graphicx}
\usepackage{dcolumn}
\usepackage{bm}

\begin{document}
\title{Geometric mechanics of random kirigami}
\author{Gaurav Chaudhary}
\affiliation{School of Engineering and Applied Sciences, Harvard University, Cambridge, MA 02138.}
\author{Lauren Niu}
\affiliation{Department of Physics, Harvard University, Cambridge, MA 02143.}
\author{Marta Lewicka}
\affiliation{Department of Mathematics, University of Pittsburgh, Pittsburgh, PA 15260.}
\author{Qing Han}
\affiliation{Department of Mathematics, University of Notre Dame, South Bend, IN 46556.}
\author{L Mahadevan}
\email{lmahadev@g.harvard.edu}
\affiliation{School of Engineering and Applied Sciences, Harvard University, Cambridge, MA 02138.}
\affiliation{Department of Physics, Harvard University, Cambridge, MA 02143.}

\begin{abstract}
The presence of cuts in a thin planar sheet can dramatically alter its mechanical and geometrical response to loading, as the cuts allow the sheet to deform strongly in the third dimension. We use  numerical experiments to characterize the geometric mechanics of kirigamized sheets as a function of the number, size and orientation of cuts. We show that the geometry of mechanically loaded sheets can be approximated as a composition of simple developable units: flats, cylinders, cones and compressed Elasticae. This geometric construction yields simple scaling laws for the mechanical response of the sheet in both the weak and strongly deformed limit. In the ultimately stretched limit, this further leads to a theorem on the nature and form of geodesics in an arbitrary kirigami pattern, consistent with observations and simulations. By varying the shape and size of the geodesic in a kirigamized sheet, we show that we can control the deployment trajectory of the sheet, and thence its functional properties as a robotic gripper or a soft light window. Overall our study of random kirigami sets the stage for controlling the shape and shielding the stresses in thin sheets using cuts.
\end{abstract}

\pacs{Valid PACS appear here}
\maketitle

Kirigami, the art of paper cutting, is now increasingly being seen as a paradigm for the design of mechanical metamaterials that exhibit exceptional geometric and structural properties \cite{zhang2015mechanically, blees2015graphene, bertoldi2017flexible}. The basis for kirigami is the well known observation that the mechanical response of thin sheets is exclusively due to the geometrical scale separation induced by slenderness which makes bending deformations inexpensive compared to stretching. In kirigami, the presence of cuts provides for an extra degree of control via the internal localization of large bending deformations at cuts; these allow for the nature and scale of internal large-scale bending modes by varying the number, size and location of the cuts. This raises a number of questions associated with both the forward problem of understanding the mechanics of these topologically and geometrically complex materials as well as the inverse problem of designing the cuts to obtain different types of articulated deformations for shape optimization. Recent work in the context of the forward problem  has focused primarily on the mechanics of kirigami with simple distributions of periodic cuts, aimed at characterizing the response using a combination of theory, experiment and computation \cite{rafsanjani2017buckling,moshe2019kirigami,sadik2021local}. In contrast, the inverse problem of designing cuts that allow for articulated shape transformations has been limited primarily to geometric optimization \cite{Choi2019,Choi2020}, without much in the nature of the mechanical response of the resulting structures. To design cut patterns to control the shape and response of kirigamized sheets, we need to combine aspects of both these classes of problems by understanding the geometric mechanics of sheets with multiple, aperiodic cuts. Here, we take a step in this direction by describing the geometry and mechanics of kirigami sheets with aperiodic, randomly located cuts in the dilute limit, so that the cuts do not themselves intersect.

\section*{Geometry and mechanics of a sheet with a single cut}

\begin{figure}[!]
	\centering
	\includegraphics[width=9 cm]{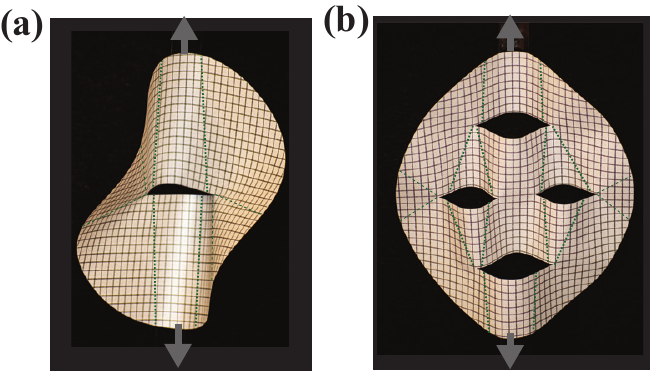}
	\caption{(a) When a thin circular paper sheet ($R=8.9$~cm, $h=0.01$~cm) with a cut of length $ l= 6.4$~cm) is loaded along diametrically opposite ends, it buckles into a complex 3D surface, but can be approximated as a composition of two cones (black dashed lines), and two cylindrical cores (in dotted lines). (b) A perforated sheet with the same dimensions as in (a) with several cuts deforms into an even more beautiful 3D structure when pulled as shown.}
	\label{fig:figure_0}
\end{figure}

To get a sense of the geometry of a kirigamized sheet, in Fig.~\ref{fig:figure_0}~a, we show the shape of a thin circular sheet of radius $R$, thickness $h$ ($R/h \gg 1$) and a single horizontal cut of length $l$. When the nominal strain  $\gamma> 0$ induced by the applied vertical force crosses a threshold, the initially planar sheet buckles out of the plane into a complex geometrical shape. If the sheet has multiple cuts, as in Fig.~\ref{fig:figure_0}~b, the deformed geometry is even more complex. However, casual observations of the sheet show that the underlying constituents of the deformed sheet are actually simple  conical and cylindrical domains connected by transition layers; increasing the topological complexity of the sheet increases the number, size, shape and orientational order of such domains. 


\begin{figure*}[htb!]
	\centering
	\includegraphics[width=0.95\textwidth]{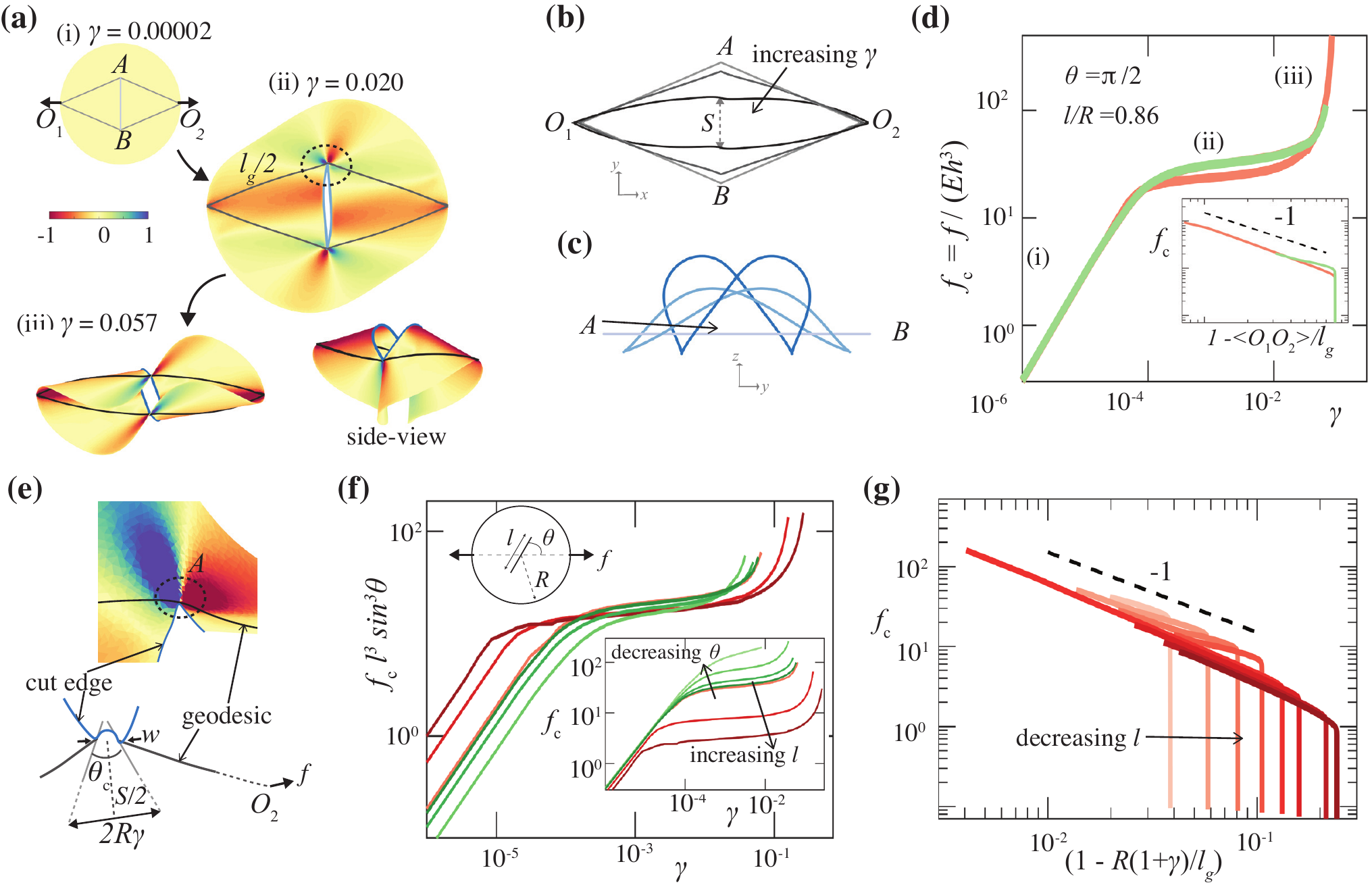}
	\caption{(a) The mean curvature of a loaded sheet with two point loads at $O_1$ and $O_2$ for various applied strains $\gamma$. The grey lines are the geodesics connecting the loading points without intersecting the cuts, and the blue lines are tracking the cut shape. Highly localized bending deformation, as indicated by the mean curvature, can be observed. (b) Geodesic connecting the loading points $O_1$, $O_2$ (extracted from Fig.~\ref{fig:figure_1}~a) better align with the axis $O_1O_2$ with increasing strain. (c) The evolving shape of the cut shows resemblance to Euler's elastica.
	(d) Force displacement curves for the case shown in (a) with $l/R = 0.86$ and $\theta = \pi/2$ subjected to point load at the opposite ends as shown in the inset. The curves show three distinct regimes corresponding to the cases shown in Fig.~\ref{fig:figure_1}~a. Green curve corresponds to the case shown in Fig.~\ref{fig:figure_1}~a and red curve corresponds to the case with cut edges buckling out-of-plane in opposite direction. The rescaled strain in a form similar to ``finite extensibiltiy" models shows a power-law scaling with an exponent $-1$. (e) A zoomed in view of the cut corner for the case Fig.~\ref{fig:figure_1}~a~(iii), and a schematic showing the relevant length scales that govern the mechanics in this regime. (f) Scaling force with ($1/(l^3 \sin^3{\theta})$) collapses the plateau mechanical response for the cases with varying cut length $l$ and orientation $\theta$. Unscaled data shown in the inset. Red curves correspond to the cases with $\theta = \pi/2$, and $l/R = 0.86, 1.29, 1.71$. Green curves are the cases with $\theta = \pi/4, \pi/3, 5\pi/12$ and $l/R = 0.86$. (d) Strongly strained kirigami sheets qualitatively show a ``freely-jointed" polymer-chain like divergence as the length $O_1 O_2 \to l_g$. Shown are the cases with $\theta = \pi/2$ and $l/R -[0.57- 1.71]$.}
	\label{fig:figure_1}
\end{figure*}
{We use numerical simulations to study the geometrical mechanics of kirigamized sheets (see SI~S2), starting from the Föppl-von Kármán plate energy of a triangulated plate that is minimized using a conjugate gradient method. Cuts are defined as thin rectangular slits of length $l$, and small width $w$, with a semicircular tip of diameter $w$ added to the tip of the cuts of width $w$.}	

Figure~\ref{fig:figure_1}~a shows the mean curvature of the deformed sheet for different values of the strain $\gamma$ when the cut is initially orthogonal ($\theta = \pi/2$) to the loading axis (the line connecting the loading points) (Fig.~\ref{fig:figure_1}~a~(i)). It is evident that for very low strains $\gamma \ll 1$, the sheet just stretches, but remains planar. When $\gamma > \gamma_c$, the sheet buckles with bending deformations becoming localized along two conical domains centered near the ends of the cut, and a large cylindrical domain of nearly uniform mean curvature $\kappa$ appears to connect the cut edges and the loading points on both sides. Further stretching of the sheet causes an increase in the curvature of these cylindrical domains, and reduces the Euclidean distance between the end-points ($A$ and $B$) of the cuts to $S$ from its initial length~$l$ (see SI Movie~S1). This deformation localization enables the straightening (in three dimensions) of the lines connecting the ends of the cuts to the points of force application $O_1, O_2$ shown in Fig.~\ref{fig:figure_1}~a, as shown in Fig.~\ref{fig:figure_1}~b. Simultaneously, the free edges associated with the cut deform into a shape that resembles Euler's elastica \cite{love2013treatise} as shown in the Fig.~\ref{fig:figure_1}~c. When the applied strain becomes very large, the end points of these elasticae, corresponding to the ends of the cut, come together so that the polygons connecting the ends of the cut to the points of force application straighten out and merge. If the sheet thickness is very small ($h/R \rightarrow 0$), we thus expect that the sheet will be locally flat-folded onto itself, a limit that we will address later. 

Moving from this geometric description of the sheet to its mechanical response, the force-displacement response of the kirigami sheet described in Fig.~\ref{fig:figure_1}~a-c is shown in Fig.~\ref{fig:figure_1}~d. At very low values of strain $\gamma$, the force is linearly proportional to the strain, but once the sheet buckles, the force flattens out to a plateau-like regime as the sheet stretches by bending out the plane. Eventually, as the ends of the cut come together, the sheet stiffens as it cannot deform any further without significant stretching and the force increases showing a near-divergence.   When the sheet buckles out of the plane, there are two almost equivalent modes of deformation: a symmetric mode when both cylindrical domains on either side of the cut are in-phase, and an antisymmetric mode both cylindrical domains on either side of the cut are out-of-phase. The plateau force for these two cases differ marginally, but the linear and the divergent response away from the plateau is indistinguishable as can be seen in Fig.~\ref{fig:figure_1}~d. 

To understand the origin of this divergence, we use a scaling approach. The torque due to the applied force $f$ acting over a length $S$ corresponding to the (small) distance between the ends of the cut is balanced by the internal elastic torque $EI \kappa_c$ where $\kappa_c$ is the characteristic mean curvature in the neighborhood of the end of the cut,  with $EI$ being the bending stiffness of the sheet ($E$ is the Youngs's modulus and $I$ is the second moment of area around the axis normal to the deformed area near the cut corner), so that $fS \sim EI \kappa_c$. As shown in Fig.~\ref{fig:figure_1}~e, at the ends of the cut of small width  $w$, the characteristic curvature $\kappa_c \sim \theta_c/w$, where $\theta_c$ is the angle at the slit corner. As the sheet is pulled apart by the forces so that they are $2R\gamma$ apart, geometry implies that $\theta_c S/2 \sim 2R \gamma$ and $S \approx 2\sqrt{(l/2)^2 - 2 R^2 \gamma}$. Substituting these geometric relations into the overall torque balance then yields the relation
\begin{equation}
    f  \sim  \frac{EI}{w R} \frac{\gamma}{(\bar{l})^2 -  \gamma} 
    \label{eq:divergence}
\end{equation}
where $\bar{l} = l/(2\sqrt{2}R)$. Writing the stretch ratio of deformation as the end-to-end displacement $R(1+\gamma)$ normalized by the length of the shortest segment connecting the force application points to the ends of the cut, i.e. the piecewise linear geodesic length $l_g = \sqrt{(l/2)^2 + R^2}$, eq.~\ref{eq:divergence} can be expressed in a more familiar form $\hat f \sim 1/(1- R(1+\gamma)/l_g)$ where $\hat f = f/(EIwR)$ which we see is similar to the divergent response of a freely-jointed polymer chain \cite{Grosberg}. The inset to Fig.~\ref{fig:figure_1}~d  shows that the mechanical response with the rescaled definition of the stretch agrees remarkably well with this simple scaling estimate. We pause to note that  the divergent mechanical response of freely-jointed chain is intimately linked to the balance between entropic effects and a finite chain length, quite unlike the divergent response of the athermal kirigami sheet which is due to the localization of curvature of the sheet at the ends of the cut. 

Having understood the geometry and mechanical response of a sheet with a single symmetrically placed cut, we ask what would happen when the length $l$ and/or its orientation $\theta$ is varied. Fig.~\ref{fig:figure_1}~f inset shows the mechanical response for various cases where $l$ is varied keeping the cut orientation orthogonal to the clamped axis ($\theta = \pi/2$) in red curves. The mechanical response in all cases is qualitatively similar to Fig.~\ref{fig:figure_1}~d, with a shift in the applied strain $\gamma$ at the onset of plateau response, the magnitude of the plateau force, and the strain at the onset of divergent response. Larger $l$ result in a lower value of the threshold in $\gamma$ and a lower force plateau persisting for longer, before the force diverges. 
Similarly changing the orientation $\theta$ of the cut changes the mechanical response and the plateau value increases as the initial cut direction is more aligned with the direction of the clamping axis. The deformed geometric configurations for all the cases are shown in Fig.~S5-6.  

To quantify these observations, we note that the linear response at very small strain corresponds to the planar stretching of sheet. The presence of the cut leads to a stress intensification near the ends of the cut. Following Inglis' seminal work \cite{inglis1913stresses}, the stress near the tip of a cut of length $2l$ and a very small radius of curvature $w$ can be approximated as $\sigma_0 \sqrt{l/w}$, where $\sigma_0$ is the far field applied stress on a large plate. Assuming the planar elastic deformation is confined near the crack tip, we get $\sigma_0 \sqrt{l/w} \sim E\gamma$. Hence we expect $\sigma_0\sim f\sim 1/\sqrt{l}$ (see SI Fig.~S4b). Following the initial linear increase in the force with applied strain, the sheet buckles when the compressive load on a plate whose size scales as the size of the cut reaches the buckling threshold. Since the buckling stress $\sigma_b \sim B/hl^2$, with $B$ being the flexural rigidity of the sheet, thus  the buckling force $f_b \sim (B/h^2 R) (1/l^{2})$ (see SI Fig.~S4d). The strain at which sheet buckles ($\gamma_b$) can be estimated from a balance of the linearly increasing planar stress ($\sim E \gamma_b/\sqrt{l/w}$)  with the buckling stress, so that $\gamma_b\sim 1/l^{3/2}$ (see also SI Fig.~S5-6).

Following the onset of buckling, for small out-of-plane displacements $\delta \ll l$ of the cut boundary, the curvature of the deformed cylindrical core of the sheet scales as $\kappa \sim \delta/ l^2$. Thus the total bending energy of the sheet scales as $U \sim B\kappa^2 A \sim B\delta^2R/ l^3$, with $A \sim Rl$ being the area of the localization size approximated by area under quadrilateral $O_1 A O_2 B$ shown in Fig.~\ref{fig:figure_1}~b. Thus the force scales as $f = \partial U/\partial (R\gamma) \sim BR\delta/l^3$. For a cut oriented at an arbitrary angle $\theta$ to the clamped axis, we replace $l$ by its orthogonal projection to the loading axis $l\sin{\theta}$. Fig.~\ref{fig:figure_1}~f shows the collapsed data from Fig.~\ref{fig:figure_1}~f inset, indicating that the bending energy localization in the elasticae corresponding to the two edges of cut determines the magnitude of plateau response in the force-displacement curves. We note that we have ignored the contribution of energy from the conical domains since the size of conical domains is much smaller than the cylinder-like domains just above the onset of out-of-plane deformation, and further mean curvature decays away from the cone tip as $\kappa \sim 1/r$. At large applied strain, when the geodesics are relatively better aligned with the loading axis, application of further strain induces a strong bending deformation at the cut corners.  Sheets with a finite tearing threshold stress generally tear as a result of this deformation, and this raises a class of different questions about the nature and shape of the curve of tearing \cite{zehnder1998williams}. Fig.~\ref{fig:figure_1}~g depicts the rescaled force-displacement curves showing the divergence response of sheets with varying cut length which agree well with the scaling arguments presented above in eq.~\ref{eq:divergence}.

\section*{Geometry and mechanics of a sheet with multiple cuts}

\begin{figure*} [htb!]
	\centering
	\includegraphics[width=15.0 cm]{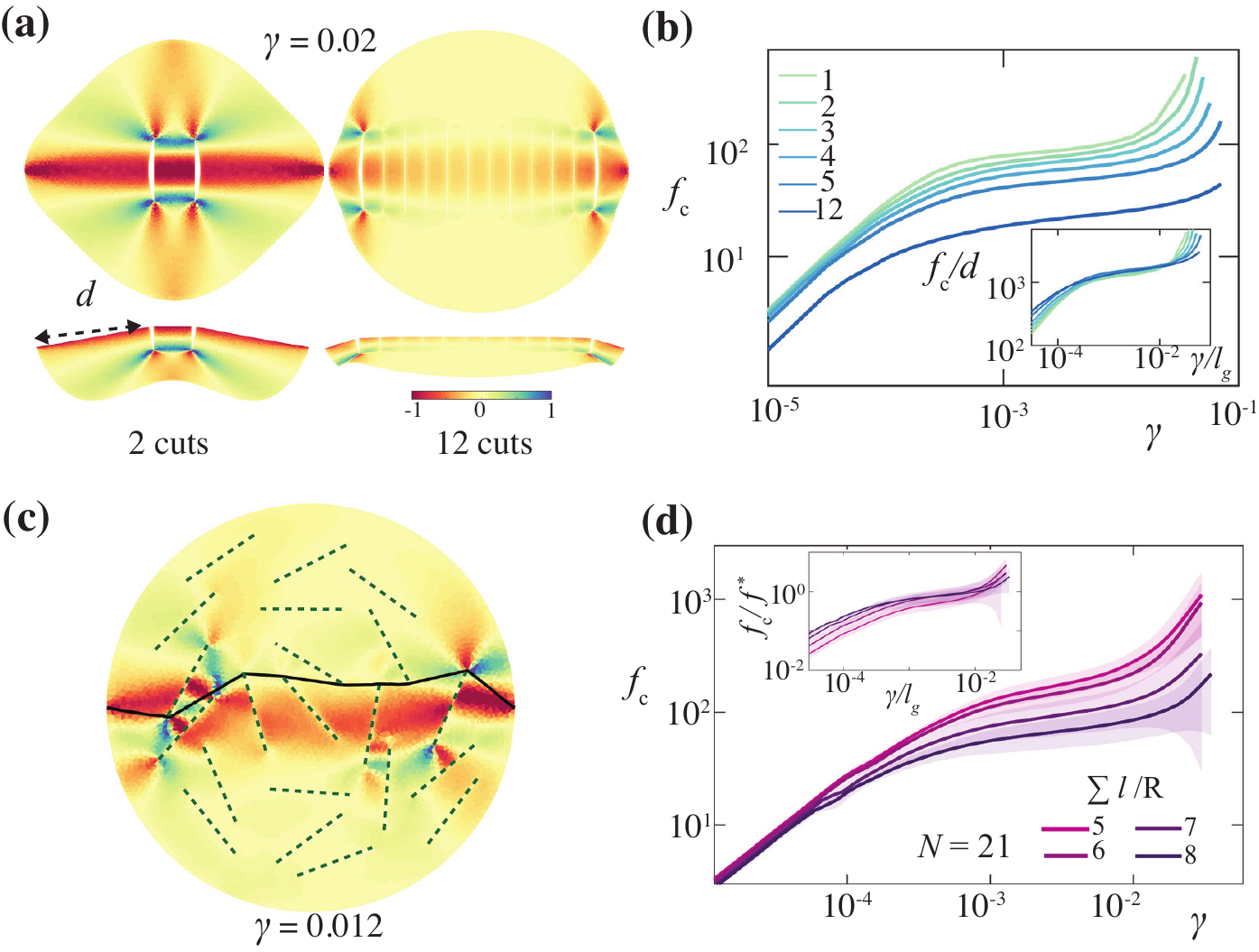}
	\caption{(a) Localized bending deformation in the sheets with multiple cuts aligned orthogonally to the loading axis show similar geometric structures regardless the number of cuts, shown are front and side views of the cases with 2 and 12 cuts. (b) Mechanical response for kirigami sheets with different number parallel of cuts ($\theta = \pi/2, l = 0.57 R$). (inset) Rescaled force with the length $d$ (labelled in (a)) collapsed the plateau regime. (c) Localized deformation in a random kirigami sheet with 21 randomly distributed cuts of same length (dashed lines) with a total length of all cuts $8R$. The mean curvature of the deformed sheet is projected on the initial flat configuration. (d) Mechanical response of a randomly kirigamized sheet perforated with 21 randomly distributed uniform cuts, with varying total cut length has similar behavior as sheets with single and multiple structured cuts. (inset) Rescaled force data collapsed the plateau force reasonably well.}
 \label{fig:figure_2}
\end{figure*}

We now turn to understand how the geometric mechanics of sheets kirigamized with a single cut translates into our understanding of sheets with multiple cuts, as shown in Fig.~\ref{fig:figure_2}. As examples, the mean curvature maps in Fig.~\ref{fig:figure_2}~a show the similarity in the localized deformation for the cases with 2 and 12 cuts that are perpendicular to the loading axis (see SI Movie S2, other cases are shown in SI Fig. S9). It is clear that the deformed geometry in such cases consists of four conical domains, and $N+1$ connected elasticae  where $N$ is the number of cuts. Indeed, irrespective of the number of cuts, the conical domains localize near the ends of the cuts that are nearest to the loading points and are connected by a flat sheet for the cases with larger $N$. Fig.~\ref{fig:figure_2}~b shows the mechanical response of sheets kirigamized with varying number of cuts and shows that increasing the number of cuts softens the system, reducing the plateau force and delaying the transition to the ultimately divergent force-displacement response. To characterize the mechanical response in the case with multiple cuts,  we approximate the bending energy localized in $N+1$ elasticae close to the onset of out-of-plane deformation.   This results in the form the energy $U \sim B R^2 d\gamma/l^3$, and hence $f \sim \partial U/\partial(R \gamma)\sim BRd/l^3$  where $d$ is the distance between the loading point and the nearest cut (see SI-S3). Consistent with this, rescaling the force with $d$ collapses the data over the scale of intermediate deformations as shown in the inset of Fig.~\ref{fig:figure_2}~b. 

At large strains the strong bending deformations near the cone tips similar to the case with single cut discussed earlier.   Predicting the location of conical domains will enable predicting the stress concentrations, and the potential sites of structural failure in practical applications. Our observations  suggest that the conical domains appear at the end of a cut if the addition of that cut increases the geodesic length. To formalize this, we define a binary participation ratio (\emph{PR}) for each cut as
\[
PR_i =
\begin{cases}
1  & \text{if $l_g^0-l_g^i >0$,} \\
0 & \text{if $l_g^0-l_g^i =0$} 
\end{cases}
\]
where $l_g^0$ is the geodesic length evaluated for a given cut arrangement, and $l_g^i$ is the geodesic length with the $i^{\text{th}}$ cut removed from the arrangement. For the cases shown in Fig.~\ref{fig:figure_2}~a-b, two cuts that are nearest to the loading points have a $PR=1$ and all other cuts have $PR=0$. For such cases, any cut with a projected length $l \sin \theta > d$ the distance of the force application point does not influence $l_g$, and thence the mechanical response. Similarly, for the cases with two cuts of varying projected lengths (see SI~ Fig. S7, S8), conical domains disappear at the corners of cuts with $PR=0$. In Fig.~\ref{fig:figure_2}~a-b, we see that an increase in the number of cuts increases the length $l_g$. Since the divergent force-displacement response emerges as $2R(1+\gamma)\to l_g$, the plateau response in the force-displacement curves is observed at larger $\gamma$ for larger $N$. In cases with $N>2$, fixing the location of the cuts closest to the points of force application sets the trajectory of geodesics as well as $l_g$, regardless the presence of the inner cuts; hence the divergence transition occurs at same strain regardless of the presence of the inner cuts. In fact the force-displacement curves overlap at all applied strains for such cases, and the geometry of the deformed sheets is identical (see SI Fig. S10). 

The geometry of the geodesics controls the mechanical response of the overall system when the cuts are randomly distributed. In Fig.~\ref{fig:figure_2}~c, we show an example with 21 cuts of the same length $8R$, with the location of the cut midpoint and its orientation randomly chosen such that a minimum separation exists between the cuts, and from the clamped points (varying the cut length does not change any of our results qualitatively). The localization of deformation in multiple elasticae and conical domains is evident while the sheet remains flat and undeformed near some cuts.

To obtain the average response for the random kirigami cases, we repeat the simulations keeping $\sum l$, sum of all cut lengths, a constant. Fig.~\ref{fig:figure_2}~d shows the mechanical response for random kirigami. The results represent mean statistics of the response for 10 samples per case. The overall nature of the force-displacement curve is similar to that of a single cut, and as expected the plateau force decreases for the cases with longer cuts. The cases with smaller $\sum l$ show a very weak deviation from the initial linear response, but with increasing $\sum l$ a clear plateau is observed. The force-displacement divergence is observed when the geodesic connecting the points of force application straightens out under applied strain.   Relatively large variance in force beyond the initial linear response exists.  This is due to the strong dependence of mechanics on the cut length and location beyond the initial linear regime. Similarly, increasing the number of cuts while keeping $\sum l$ constant, results in similar observations (see SI~Fig.~S12) with a lower number of longer length cuts resulting in a lower plateau force response and an increased variance. 

It is evident from the geometry of random kirigami that the deformation gets localized near a few cuts (see SI Fig.~S11). Just as for the case with structured cuts, where the mechanical response depends on the distance of cut from the point of application of force, its projected length and its participation ratio $PR$, for random cuts that are not very close to one another, a similar scenario arises. At the onset of the buckling transition from the initial planar stretching response at very low strains, each cut with $PR>0$ introduces a soft bending deformation mode in the sheet with a characteristic bending force $f^*$, given by the smallest buckling load, i.e. 
\begin{equation}
  f^* \sim Eh^3 ~\min_i \left[\frac{d_i}{l_i^3\sin^3{\theta_{d_i}}} +    \frac{2R-d_i}{l_i^3\sin^3{\theta_{2R-d_i}}}\right]2R
  \label{eq:characteristic_force}
\end{equation}
where $d_i$ is the minimum distance of the midpoint of the cut of length $l_i$ from the loading points, and $2R-d_i$ is the distance of the cut midpoint from the farthest loading point, and $\theta_{d_i}$ and $\theta_{2R-d_i}$ are the angles that the cut makes with the line joining the midpoint of the cut to the points of force application. The above result follows from the assumption that at the onset of plateau regime, the characteristic mean curvature for bending localization is set by the cut that corresponds to $f^*$, so that the two terms in eq.~\ref{eq:characteristic_force} follow from the energy of two elasticae that exist on both sides of the cut. Here, we note that the cuts with $PR=0$ do not alter the plateau response near its onset as seen with the cases in Fig.~\ref{fig:figure_2}~b (see SI~Fig.~S7-8). These observations allow us to determine the rescaled mechanical response shown in the inset of Fig.~\ref{fig:figure_2}~d, providing a reasonable collapse in the plateau region of force-displacement data. The spread in the scaled data is likely due to the simplification that the area of the sheet where energy is localized is assumed to span the sheet (hence the factor $2R$ in eq.~\ref{eq:characteristic_force}), and that additional cuts which buckle following the onset of first buckling also contribute to localize the bending deformation. All together, this allows us to reduce a given random kirigamized sheet to a simpler, \lq\lq mechanical equivalent\rq\rq ~(see SI Fig.~S13). Since the cuts with $PR=0$ do not effect the geometric mechanics of the sheet, simply removing them from the given cut arrangement results in a sheets with identically geometric mechanics with reduced number of cuts. Thus our simple scaling approach reduces the complexity of random kirigami using elementary geometric mechanics. 

In the ultimate deformation limit of sheets with multiple cuts, the shape of the sheet is determined by the (3-dimensional) straightening of the shortest path (corresponding to the geodesic) connecting the points of force application. Geodesics for all the  cases have three components: two line segments connecting the points of application of force to the ends of the cuts nearest to these points, and a piecewise polygonal geodesic passing through the ends of all cuts.

\section*{Zero-thickness, flat-folded kirigami}	

\begin{figure*} [htb!]
	\centering
	\includegraphics[width=14 cm]{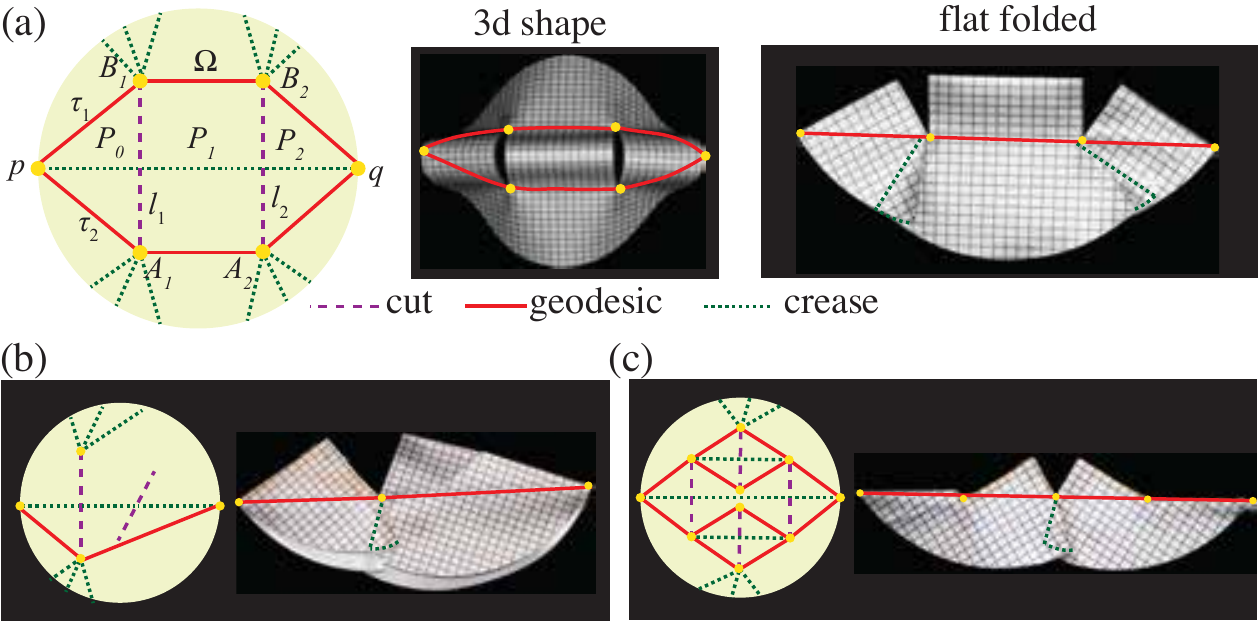}
	\caption{Strongly stretched inextensible sheet of negligible bending rigidity, $h\to 0$, can be folded to 2d sheets. This is illustrated through three different kirigamized sheets. The two projections are experimentally realized by creasing the sheets: (i) 1 crease along each cylindrical core, and (ii) 3 creases to flatten the conical domains near the cut corners. The flat foldability of a strongly stretched sheet leads to straightening of geodesics (shown in red curve) in 3d space.}
	\label{fig:figure_proof}
	\end{figure*}

For an initially flat sheet, our observations suggest that the shortest path between the points of force application for simple cut patterns is just a polygonal curve that connects these points, i.e. all geodesics in a planar sheet with random cuts are polygonals. When a very thin sheet is deformed by boundary forces, its ultimate shape is characterized by the formation of sharp creases as the sheet folds on itself, as shown in Fig.~\ref{fig:figure_proof}. These observations of the geometry of strongly deformed kirigamized sheets show that the polygonal geodesic connecting the points of force application in the plane, becomes approximately straight $\mathbb{R}^3$. When the sheet is flat-folded, the geodesic is rectified leading to a configuration that is a piecewise affine isometric immersion of the plane. We leave precise theorems and proofs of these statements for a separate study \cite{HLM2021}, but provide an intuitive argument for them here.

\smallskip

\begin{figure}[h!]
\centering
\includegraphics[scale=0.4]{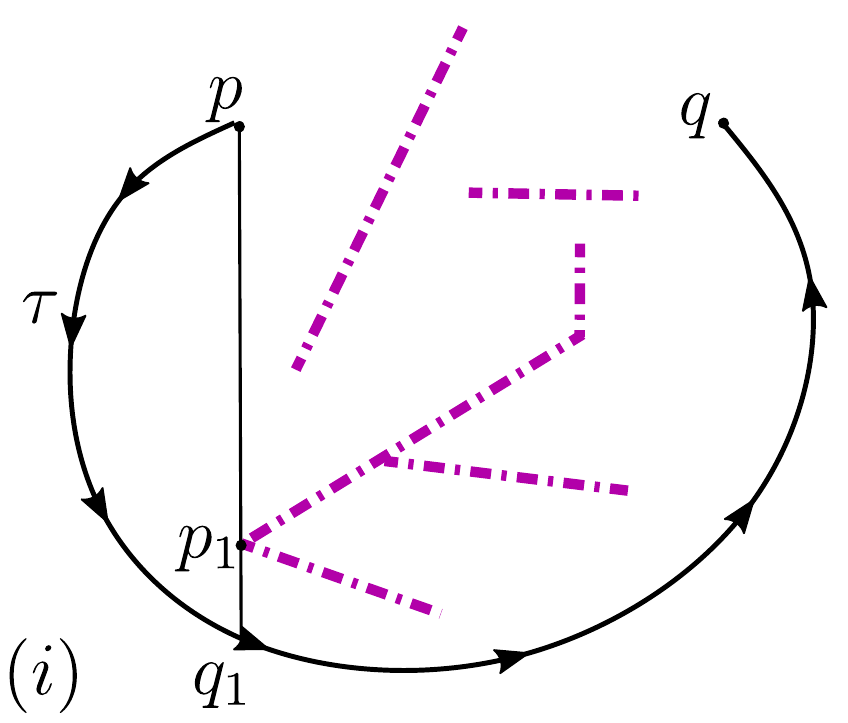}\qquad 
\quad \includegraphics[scale=0.4]{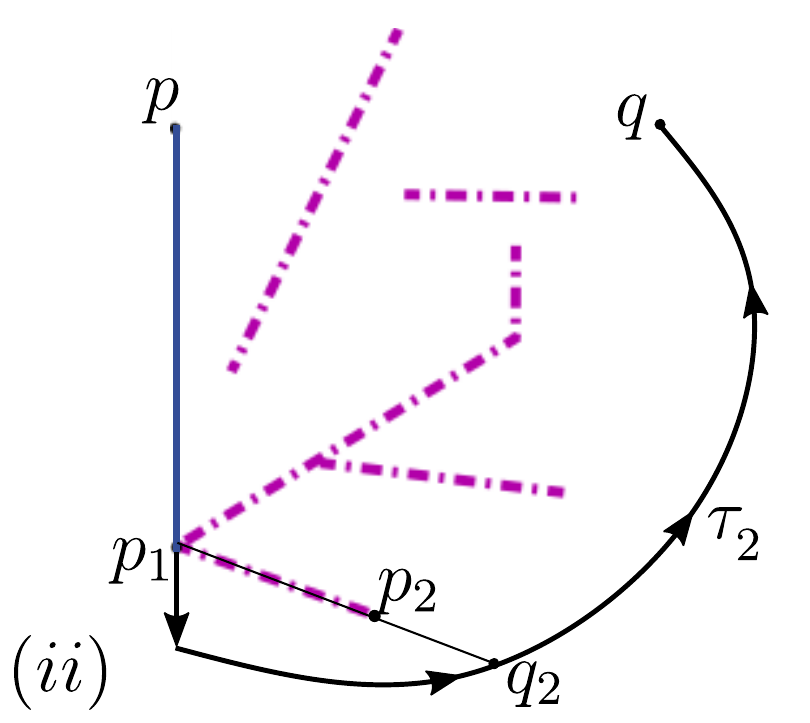} \vspace{0.8cm} \\
 \includegraphics[scale=0.4]{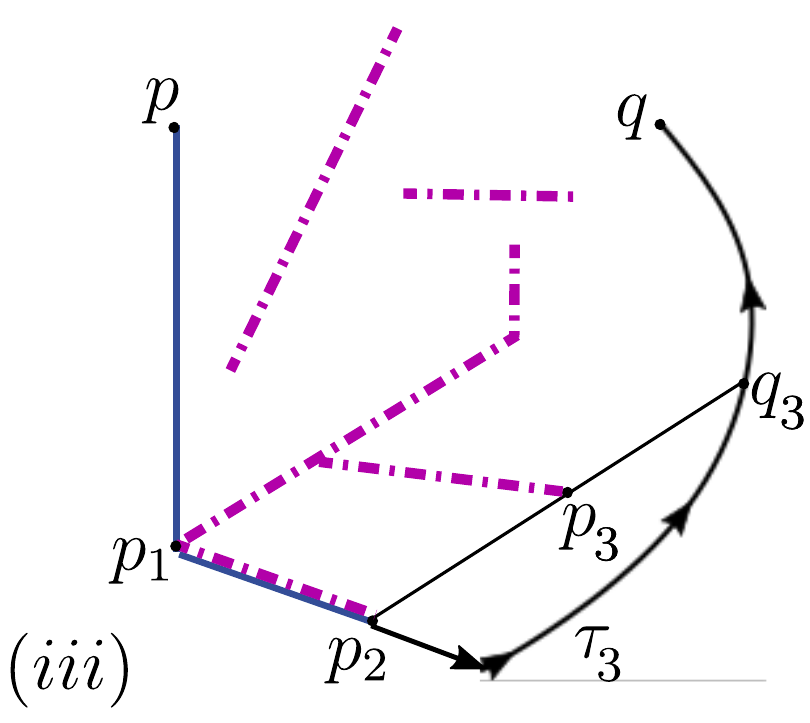} \qquad 
\quad \includegraphics[scale=0.4]{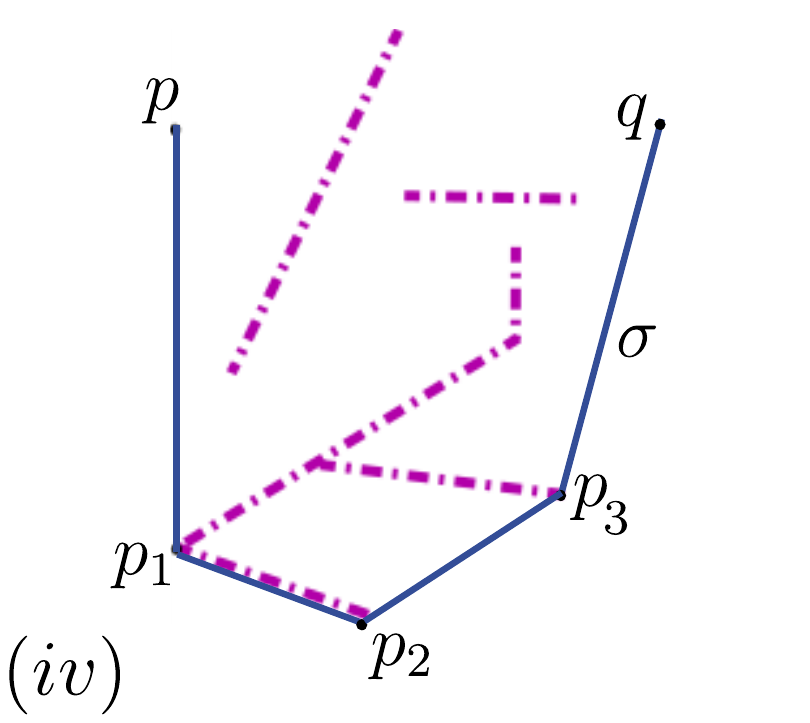}
\caption{{The path-shortening algorithm yields a polygonal competing to be a geodesic.}}
\label{Fig_algorithm}
\end{figure}	

We represent the given set of cuts $L$ contained in an open, bounded, convex domain $\Omega\subset \mathbb{R}^2$, as the set of edges of a graph $G$. Without loss of generality $G$ is  planar,
i.e. each pair of its edges intersects at most at a single common vertex.

The first result, i.e. the polygonal structure of geodesics, is shown via a path-shortening algorithm. Given $p, q\in \bar\Omega\setminus L$ and a piecewise $C^1$ curve $\tau: [0, 1]\to\bar\Omega\setminus L$ with $\tau(0) = p$ and $\tau(1) = q$, one successively replaces its portions by segments, as follows. 
Let $t_1\in (0,1]$ be the first time that the segment $\overline{p\tau(t_1)}$ intersects $L$. 
If $\tau(t_1)=q$, then $\overline{pq}$ is the desired geodesic connecting $p$ and $q$. 
Otherwise, $\overline{p\tau(t_1)}$ must contain some of
the vertices of the cuts. Call $p_1$ the closest one of these vertices to $\tau(t_1)$ and concatenate the segment
$\overline{p_1\tau(t_1)}$ with the curve $\tau$ restricted to $[t_1, 1]$. The process is now repeated from $p_1$. After finitely many such
steps one obtains a polygonal connecting $p$ and $q$,
with length shorter length than that of $\tau$.

The second result, i.e. existence of a geodesic-rectifying piecewise affine isometry, follows via the folding algorithm below. We take $p, q\in\partial\Omega$ and denote by $d$
the length of the geodesics between $p$ and $q$. 

{\it Step 1. Sealing portions of {\it inessential cuts} that do not affect $\mbox{dist}(p, q)$.} 
To this end, label cuts (the edges of $G$) by $l_1, \ldots, l_{n}$. 
Move the first endpoint vertex of $l_1$ toward its second vertex, and
start \lq\lq sealing\rq\rq\ the portion of the cut $l_1$ left
behind. The length $d$ of the geodesics connecting $p$ and $q$ is nonincreasing: it may drop
initially, decrease continuously, or it may initially remain constant. The sealing
process is stopped when $d$ becomes strictly
less than the original geodesic distance, and the new position point is labeled as
the new vertex endpoint of $l_1$. In the next step, the second
endpoint is moved along $l_1$ toward the (new) first endpoint and the
process is repeated, thus possibly sealing the cut $l_1$ further. The
same procedure is carried out for each $l_i$ in the given order  $i =
1, \ldots, n$. It follows that upon repeating the same process
for the newly created configuration, labeled the {\it minimal configuration}, it will not be further altered. 

While different ordering of cuts and vertices may yield different minimal configurations and new geodesics may be created in the cut-sealing process, all original geodesics are preserved. Also, since the updated set $L$ is a subset of the original $L$, finding an isometry relative to the new $L$ yields an isometry for the original cut set $L$.

{\it Step 2: Ordering the geodesics and the connected components of $\Omega\setminus L$.} 
There are two important properties of any minimal configuration:  the graph $G$ has no loops (i.e. it is a collection of its connected
components that are trees), and each vertex that is a leaf is a vertex of some geodesic. 

With these properties, one proceeds to label all geodesics in a consecutive order, with $\tau_1\preceq\ldots\preceq\tau_N$. Here, $\tau_r\preceq\tau_{r+1}$ means that the concatenated polygonal from $p$ to $q$ via $\tau_r$ and then back to $p$ via $\tau_{r+1}$ encloses a region $D_r$ and it is oriented counterclockwise with respect to $D_r$. Next, one labels and orders the trees $\{T_m\}_{m=1}^s$ in $\bar D_r$ so that $D_r\setminus L$ is partitioned into subregions $\{P_m\}_{m=0}^s$ and $\{Q_m\}_{m=1}^s$ in the following way: each $P_m$ is a polygon bounded by the \lq\lq  right most\rq\rq\ path from the tree $T_{m}$, the \lq\lq left most\rq\rq\ path from $T_{m+1}$,   and the intermediate portions of $\tau_r$ and $\tau_{r+1}$ which are concave with respect to $P_{m}$. Each $Q_m$ is a finite union of polygons enclosed within the single tree $T_m$, again bounded by portions of geodesics $\tau_r$ and $\tau_{r+1}$. 
Note that $\tau_r$ and $\tau_{r+1}$ may have nontrivial overlaps and some of $\{Q_m\}_{m=1}^s$ may be empty.

{\it Step 3: Constructing a desired isometry.}
We fix the segment $I=[0,de_1]$ along the $x_1$-axis in $\mathbb R^3$ and construct an isometric immersion $u$ of $\Omega\setminus L$ into $R^3$, such that $u(p)=0$, $u(q)=de_1$, $u(\tau_r)=I$ for $r=1,\ldots, N$, and where
each segment on $\tau_r$ is mapped onto a designated subsegment of $I$. 

The resulting $u$ consists exclusively of planar folds and returns the image that is a subset of $\mathbb R^2$.
By {\it Step 2}, for each $r=1, \cdots, N-1$ we have $D_r=\bigcup_{m=0}^s P_{m}\cup \bigcup_{m=1}^s Q_m$.
We construct $u$ separately on $P_0, Q_1, Q_2, \ldots, Q_m,
P_m$, where the step to construct $u$ on $P_1, \ldots, P_{m-1}$ is highly technical \cite{HLM2021}. Since the exterior region $D_0=\Omega\setminus\bigcup_{r=0}^{N-1}D_r$ does not contain trees, the two outermost geodesics $\sigma_1$ and $\sigma_N$ are convex, and so the definition of $u$ on $D_0$ consists of several simple folds. 

It also turns out that the condition $p,q\in\partial\Omega$ is essential: there exist minimal configurations for $p,q\notin\partial\Omega$, that do not admit any isometry $u$ with the property that the Euclidean distance from $u(p)$ to $u(q)$ equals the geodesic distance from $p$ to $q$ in $\Omega\setminus L$ (for further details, we refer to \cite{HLM2021}).  

\section*{Functional kirigami structures}

\begin{figure}[htb!]
	\centering
	\includegraphics[width= 11 cm]{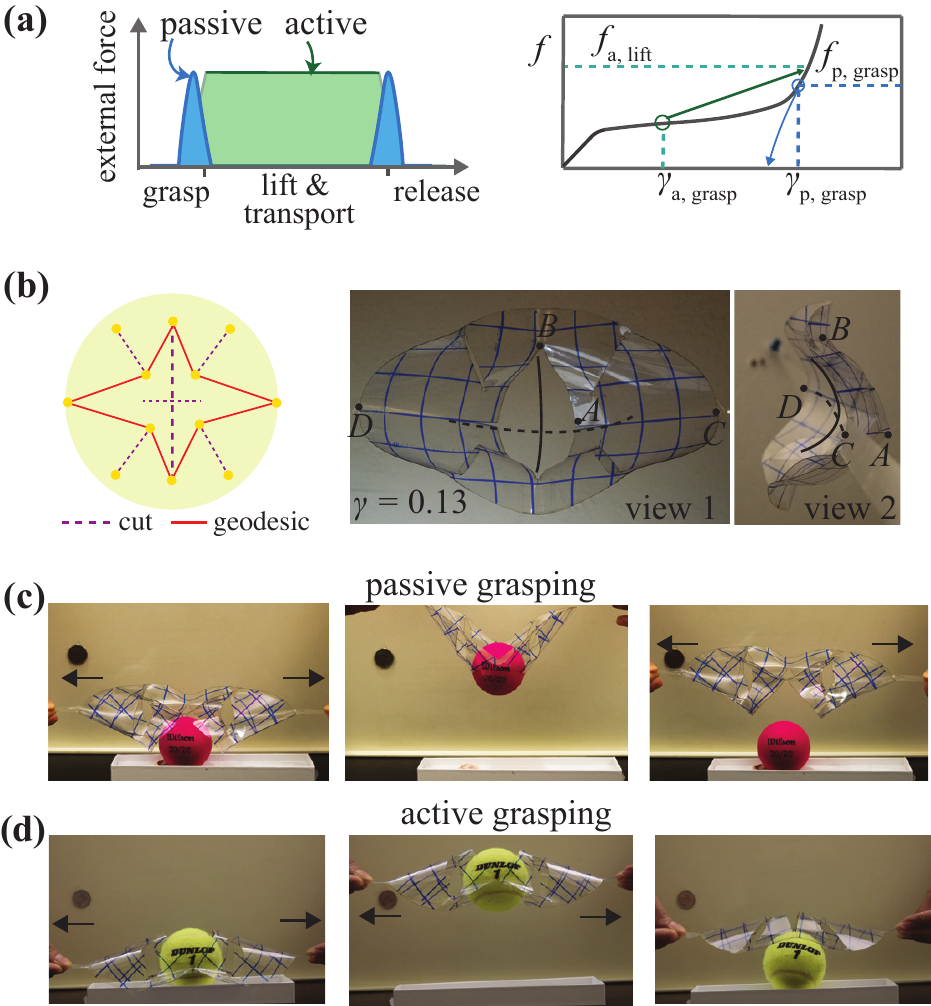}
	\caption{(a) The kirigami grasper can be employed in active and passive modes requiring different energy input. A simple framework of grasper function on a generic mechanical displacement of the sheet. (b) A kirigami grasper design for grasping objects. The deformed sheet shows the concave regions (for active grasping), and open holes (for passive grasping). (c) Passsive grasping of an racquet ball. (d) Active grasping of a tennis ball.}
	\label{fig:figure_5}
\end{figure}

The geometric mechanics of ordered and disordered kirigami leads us naturally to questions of design for function, which we demonstrate with two examples. The first is the use of kirigamized sheets to robotic grasping. To be effective, a grasper must enable controlled gripping, lifting and relocating objects of varying scales and shapes with minimal external energy input. We define an active grasper as one that requires continuous application of tensile force (external energy) to grasp and relocate, while a passive grasper only requires external work to be done in order to grasp and release the object, with \emph{energy-free} relocation as shown in the schematic in Fig.~\ref{fig:figure_5}~a. Kirigami enables both designs;
while the active gripper accommodates the target object in the curved features of the deformed sheet, the passive gripper utilizes the holes/cuts in the structure to accommodate objects. The passive mode thus requires a prestretch to deform the 2d cut into a 3d slot/hole of the size similar to the target object. On releasing the prestretch the cut boundary forms contact with the object, grasping it. The arrested object is released by applying an extensional strain to the sheet. The sequence is demonstrated in Fig.~\ref{fig:figure_5}~c and SI Movie S3. In the passive mode, the conical tips and the cylindrical core of the kirigami sheet enables confinement of the target object as shown in Fig.~\ref{fig:figure_5}~c and SI Movie S3.    

The kirigamized grasper has multiple smaller cuts, in addition to the larger cut in the middle as shown in Fig.~\ref{fig:figure_5}~b. The additional cuts are prescribed in a way that all cuts have $PR>0$, and buckle under the applied strain. Further, the largest cut is split in the middle by a small cut, oriented along the pulling axis. Together these features enable symmetric deformation of the flat kirigami sheet, and improve the stability in handling due to additional points of contact with the grasped object as shown in Fig.~\ref{fig:figure_5}~b. We note that the simplest kirigamized gripper with a single cut (Fig.~\ref{fig:figure_0}~a, ~\ref{fig:figure_1}~a) deforms asymmetrically, and hence cannot be effectively used as a grasper in both active and passive modes. In practical scenarios however symmetric deformation can be realized if the sheet is stretched significantly \cite{yang2021grasping}.

The range of applicability of a kirigami gripper can be understood from a balance of the forces due to the bending of kirigami sheet, and the weight of the object to grasp. For a cut of length $l$, the characteristic grasping force of the deformed sheet can be written as $f_g \sim \mu (B/l) c(\gamma)$, with $\mu$ being the coefficient of friction and $c(\gamma)$ is a strain dependent geometric factor. For an object with effective density $\rho$ and size $l$, force balance yields $\mu (B/l) c(\gamma) \sim \rho g l^3$. This provides a non-dimensional kirigami grasper parameter $\rho g l^4/(\mu B c(\gamma)) $, which has values ranging between $1-100$ for the successful grasping demonstrations using a plastic gripper.

Although our discussion so far was restricted to cuts that are straight rectangular slits, the ideas established are applicable to general shaped cuts. Inspired by the recent efforts towards design of kirigami-inspired mechanically deployable structures \cite{zhang2015mechanically}, we demonstrate a simple deployable kirigami structure here, whose force-dependent shielding and transmission can be tuned with simple geometric parameters.

\begin{figure}[t!]
	\centering
	\includegraphics[width= 11 cm]{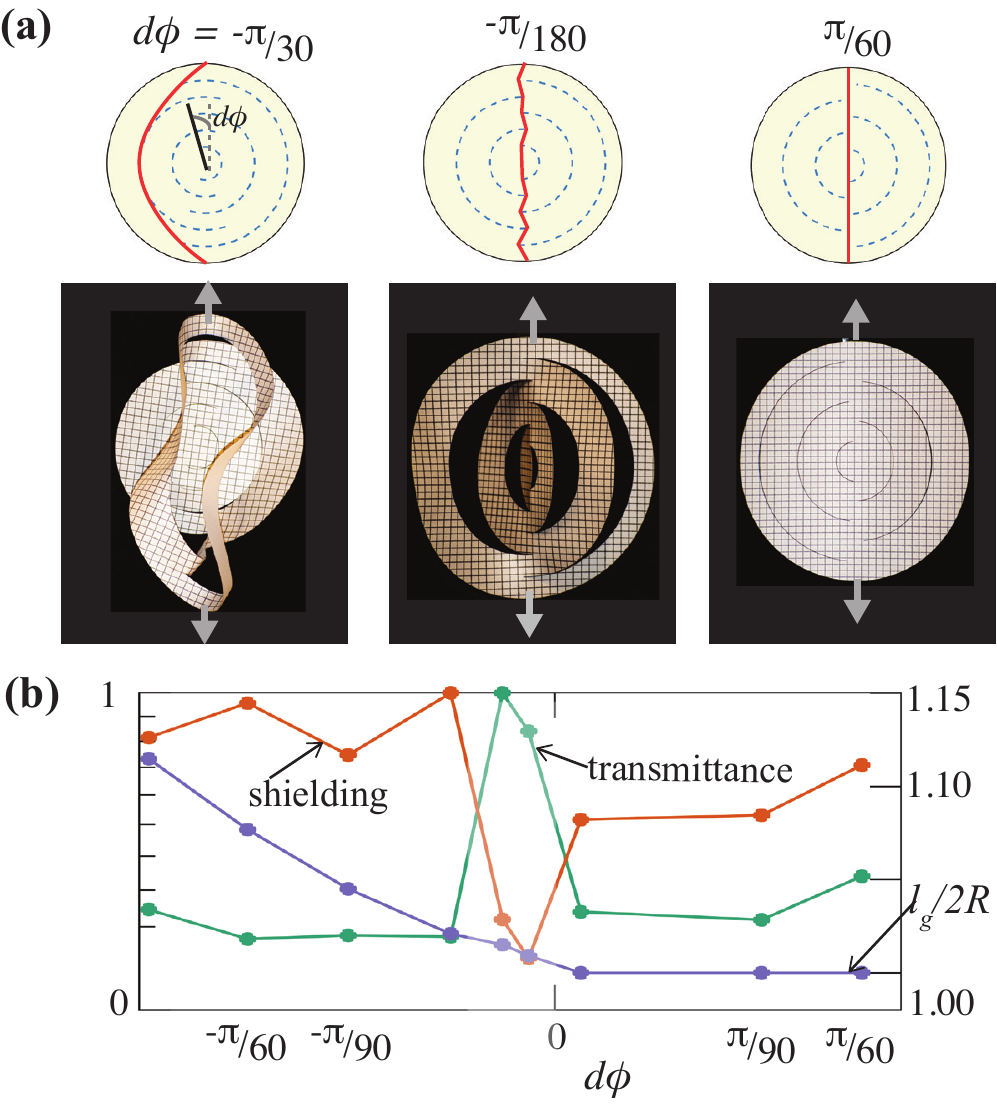}
	\caption{(a) Kirigami sheets with curved cuts (dashed lines) show very different response to an applied strain. Red line is the geodesic connecting points of application of force $O_1, O_2$. (b) A phase diagram quantifying the mechanical shielding and transmittance response of the deployed kirigami structures.}
	\label{fig:figure_6}
\end{figure}

Fig.~\ref{fig:figure_6}~a shows a case of physical kirigamized sheets perforated with five concentric circular arcs. The arcs extend between $\pm d\phi$ and $\pm(\pi - d\phi)$ w.r.t. the vertical axis, and the radius is linearly increased between the five arcs. The planar kirigami sheets (with marginally different $d\phi$) show very different geometric mechanics under a small applied strain. For the case when $ d\phi = -\pi/30$, the geodesic connecting $O_1$ and $O_2$ skirts the outer cut without intersecting any of the inner cuts. The inner cuts have $PR=0$, and hence do not introduce any soft deformation modes. Under an applied deformation, the outer frame localizes the bending while the inner structure stays nearly planar without any deformation or stays ``mechanically shielded''. A small change in $d\phi (-\pi/180)$ results in the cut arrangement such that the geodesic connecting $O_1$ and $O_2$ meanders through the corners of all cuts. Under an applied load, this structure shows a large relative out-of-plane displacement of different domains of the sheet (see SI Movie S4). Further changing  $d\phi = \pi/60$, results in a straight geodesic connecting $O_1$ and $O_2$, and hence tensile loading of this structure results in the in-plane deformation. 

We quantify the functional response of such kirigamized structures using $d\phi$ as a tunable parameter. A straightforward observation is that force required to deform such class of structures increases monotonically with $d\phi$ (see SI Fig.~S14a). The geometric consequence of different geodesic paths can be quantified in terms of two functional features of this family of kirigami structures: transmittance and shielding (\ref{sssec:materials_curved}c:materials). In practical scenarios this corresponds to the light transmitted though an optical window when illuminated with a light rays perpendicular to the rest plane. Shielding is proposed to be linked its ability to restrict the mechanical deformation to the boundary, and effectively protecting the interior. We restrict the deformation to small strains $0.01-1\%$, a practically relevant regime. And since the geometric mechanics is strain-dependent, we take a representative value for comparison corresponding to the highest applied strain.

Fig.~\ref{fig:figure_6}~b shows a phase diagram showing shielding and transmittance as a function of $d\phi$. The properties are evaluated at a strain of 1\%. Both flatness and transmittance curves display a non-monotonic trends. The cases with large $l_g$ show high shielding ability since their interior remains relatively flat, while a high transmittance is achievable in the cases with $l_g \to 2R$. Further the strain sensitivity of transmittance is enhanced with increasing $l_g/2R$ (see SI Fig.~S14b). For the cases with $l_g=2R$ a weak dependence on the functional properties can be seen since the structures deform primarily by planar stretching. 

\section*{Discussion}
Our study of random kirigami has shown how elementary geometric and energetic concepts allow us to understand the three-dimensional structure and mechanical response of kirigamized sheets. This leads to a geometric view of how cuts respond or not, along with scaling arguments for all regimes of deformation. The resulting simplicity of the framework reduces a complex nonlinear problem to  geometrical constructions, and thus eases the search for novel engineering solutions using kirigami in such instances as grasping and windowing, which we hope are just the beginning of a different way of thinking about using topological and geometrical mechanical metamaterials.

\section*{Materials and Methods}
\subsection*{Numerical Simulations}
\label{ssec:numerical_methods}
For our numerical experiments, use the finite difference scheme outlined in \cite{weischedel2012construction} and \cite{vanrees2017growth} to represent the sheets and minimize the Föppl-von Kármán plate energy on triangular meshes of the geometry, fixing the two clamped regions at various distances apart, with otherwise free boundary conditions. The Young (elastic) modulus, Poisson ratio, and thickness used to calculate the elastic energy are $1$~GPa, $0.4$, and $0.01$~cm respectively. The geodesics of the mesh were computed using an open source code \cite{peyre2011numerical} based on fast marching approach \cite{kimmel1998computing}.

\subsection*{Quantification of functional response}
\label{sssec:materials_curved}
Transmission of the kirgami window is quantified as the difference between the area of undeformed sheet and the projected area of deformed sheet on the rest plane, normalized by the undeformed sheet area. The shielding effect is quantified by the average flatness of the sheet. We define flatness as the variance in the radial correlation distribution of the average face normal of an area element with the orientation of the element at sheet center, $(\sum_i \hat{n}_0.\hat{n}(r_i) a(r_i))/A(r_i+\Delta r)$. Here, $\hat{n}_0$ is the unit normal to a infinitesimal element at $r_i\to 0$ and $\hat{n}(r_i)$ is the unit normal to an element at a distance $r_i$ from the center with an area $a(r_i)$. This quantity represents the average orientation of all elements located between $r_i$ and $r_i + \Delta r$ with the orientation of the center.

\begin{acknowledgments}
{We thank the Bertoldi lab, Jeremy A. Guillette and FAS Academic Technology at Harvard University for sharing resources that helped our experiments. The work was supported partially by NSF grants DMS-2006439 (ML),  BioMatter DMR 1922321 and MRSEC DMR 2011754 and EFRI 1830901, the Simons Foundation and the Seydoux Fund (LM).}
\end{acknowledgments}

\bibliography{kirigami_references}

\begin{thebibliography}{18}%
\makeatletter
\providecommand \@ifxundefined [1]{%
 \@ifx{#1\undefined}
}%
\providecommand \@ifnum [1]{%
 \ifnum #1\expandafter \@firstoftwo
 \else \expandafter \@secondoftwo
 \fi
}%
\providecommand \@ifx [1]{%
 \ifx #1\expandafter \@firstoftwo
 \else \expandafter \@secondoftwo
 \fi
}%
\providecommand \natexlab [1]{#1}%
\providecommand \enquote  [1]{``#1''}%
\providecommand \bibnamefont  [1]{#1}%
\providecommand \bibfnamefont [1]{#1}%
\providecommand \citenamefont [1]{#1}%
\providecommand \href@noop [0]{\@secondoftwo}%
\providecommand \href [0]{\begingroup \@sanitize@url \@href}%
\providecommand \@href[1]{\@@startlink{#1}\@@href}%
\providecommand \@@href[1]{\endgroup#1\@@endlink}%
\providecommand \@sanitize@url [0]{\catcode `\\12\catcode `\$12\catcode
  `\&12\catcode `\#12\catcode `\^12\catcode `\_12\catcode `\%12\relax}%
\providecommand \@@startlink[1]{}%
\providecommand \@@endlink[0]{}%
\providecommand \url  [0]{\begingroup\@sanitize@url \@url }%
\providecommand \@url [1]{\endgroup\@href {#1}{\urlprefix }}%
\providecommand \urlprefix  [0]{URL }%
\providecommand \Eprint [0]{\href }%
\providecommand \doibase [0]{http://dx.doi.org/}%
\providecommand \selectlanguage [0]{\@gobble}%
\providecommand \bibinfo  [0]{\@secondoftwo}%
\providecommand \bibfield  [0]{\@secondoftwo}%
\providecommand \translation [1]{[#1]}%
\providecommand \BibitemOpen [0]{}%
\providecommand \bibitemStop [0]{}%
\providecommand \bibitemNoStop [0]{.\EOS\space}%
\providecommand \EOS [0]{\spacefactor3000\relax}%
\providecommand \BibitemShut  [1]{\csname bibitem#1\endcsname}%
\let\auto@bib@innerbib\@empty
\bibitem [{\citenamefont {Zhang}\ \emph {et~al.}(2015)\citenamefont {Zhang},
  \citenamefont {Yan}, \citenamefont {Nan}, \citenamefont {Xiao}, \citenamefont
  {Liu}, \citenamefont {Luan}, \citenamefont {Fu}, \citenamefont {Wang},
  \citenamefont {Yang}, \citenamefont {Wang} \emph
  {et~al.}}]{zhang2015mechanically}%
  \BibitemOpen
  \bibfield  {author} {\bibinfo {author} {\bibfnamefont {Y.}~\bibnamefont
  {Zhang}}, \bibinfo {author} {\bibfnamefont {Z.}~\bibnamefont {Yan}}, \bibinfo
  {author} {\bibfnamefont {K.}~\bibnamefont {Nan}}, \bibinfo {author}
  {\bibfnamefont {D.}~\bibnamefont {Xiao}}, \bibinfo {author} {\bibfnamefont
  {Y.}~\bibnamefont {Liu}}, \bibinfo {author} {\bibfnamefont {H.}~\bibnamefont
  {Luan}}, \bibinfo {author} {\bibfnamefont {H.}~\bibnamefont {Fu}}, \bibinfo
  {author} {\bibfnamefont {X.}~\bibnamefont {Wang}}, \bibinfo {author}
  {\bibfnamefont {Q.}~\bibnamefont {Yang}}, \bibinfo {author} {\bibfnamefont
  {J.}~\bibnamefont {Wang}},  \emph {et~al.},\ }\href@noop {} {\bibfield
  {journal} {\bibinfo  {journal} {Proceedings of the National Academy of
  Sciences}\ }\textbf {\bibinfo {volume} {112}},\ \bibinfo {pages} {11757}
  (\bibinfo {year} {2015})}\BibitemShut {NoStop}%
\bibitem [{\citenamefont {Blees}\ \emph {et~al.}(2015)\citenamefont {Blees},
  \citenamefont {Barnard}, \citenamefont {Rose}, \citenamefont {Roberts},
  \citenamefont {McGill}, \citenamefont {Huang}, \citenamefont {Ruyack},
  \citenamefont {Kevek}, \citenamefont {Kobrin}, \citenamefont {Muller} \emph
  {et~al.}}]{blees2015graphene}%
  \BibitemOpen
  \bibfield  {author} {\bibinfo {author} {\bibfnamefont {M.~K.}\ \bibnamefont
  {Blees}}, \bibinfo {author} {\bibfnamefont {A.~W.}\ \bibnamefont {Barnard}},
  \bibinfo {author} {\bibfnamefont {P.~A.}\ \bibnamefont {Rose}}, \bibinfo
  {author} {\bibfnamefont {S.~P.}\ \bibnamefont {Roberts}}, \bibinfo {author}
  {\bibfnamefont {K.~L.}\ \bibnamefont {McGill}}, \bibinfo {author}
  {\bibfnamefont {P.~Y.}\ \bibnamefont {Huang}}, \bibinfo {author}
  {\bibfnamefont {A.~R.}\ \bibnamefont {Ruyack}}, \bibinfo {author}
  {\bibfnamefont {J.~W.}\ \bibnamefont {Kevek}}, \bibinfo {author}
  {\bibfnamefont {B.}~\bibnamefont {Kobrin}}, \bibinfo {author} {\bibfnamefont
  {D.~A.}\ \bibnamefont {Muller}},  \emph {et~al.},\ }\href@noop {} {\bibfield
  {journal} {\bibinfo  {journal} {Nature}\ }\textbf {\bibinfo {volume} {524}},\
  \bibinfo {pages} {204} (\bibinfo {year} {2015})}\BibitemShut {NoStop}%
\bibitem [{\citenamefont {Bertoldi}\ \emph {et~al.}(2017)\citenamefont
  {Bertoldi}, \citenamefont {Vitelli}, \citenamefont {Christensen},\ and\
  \citenamefont {Van~Hecke}}]{bertoldi2017flexible}%
  \BibitemOpen
  \bibfield  {author} {\bibinfo {author} {\bibfnamefont {K.}~\bibnamefont
  {Bertoldi}}, \bibinfo {author} {\bibfnamefont {V.}~\bibnamefont {Vitelli}},
  \bibinfo {author} {\bibfnamefont {J.}~\bibnamefont {Christensen}}, \ and\
  \bibinfo {author} {\bibfnamefont {M.}~\bibnamefont {Van~Hecke}},\ }\href@noop
  {} {\bibfield  {journal} {\bibinfo  {journal} {Nature Reviews Materials}\
  }\textbf {\bibinfo {volume} {2}},\ \bibinfo {pages} {1} (\bibinfo {year}
  {2017})}\BibitemShut {NoStop}%
\bibitem [{\citenamefont {Rafsanjani}\ and\ \citenamefont
  {Bertoldi}(2017)}]{rafsanjani2017buckling}%
  \BibitemOpen
  \bibfield  {author} {\bibinfo {author} {\bibfnamefont {A.}~\bibnamefont
  {Rafsanjani}}\ and\ \bibinfo {author} {\bibfnamefont {K.}~\bibnamefont
  {Bertoldi}},\ }\href@noop {} {\bibfield  {journal} {\bibinfo  {journal}
  {Physical Review Letters}\ }\textbf {\bibinfo {volume} {118}},\ \bibinfo
  {pages} {084301} (\bibinfo {year} {2017})}\BibitemShut {NoStop}%
\bibitem [{\citenamefont {Moshe}\ \emph {et~al.}(2019)\citenamefont {Moshe},
  \citenamefont {Esposito}, \citenamefont {Shankar}, \citenamefont {Bircan},
  \citenamefont {Cohen}, \citenamefont {Nelson},\ and\ \citenamefont
  {Bowick}}]{moshe2019kirigami}%
  \BibitemOpen
  \bibfield  {author} {\bibinfo {author} {\bibfnamefont {M.}~\bibnamefont
  {Moshe}}, \bibinfo {author} {\bibfnamefont {E.}~\bibnamefont {Esposito}},
  \bibinfo {author} {\bibfnamefont {S.}~\bibnamefont {Shankar}}, \bibinfo
  {author} {\bibfnamefont {B.}~\bibnamefont {Bircan}}, \bibinfo {author}
  {\bibfnamefont {I.}~\bibnamefont {Cohen}}, \bibinfo {author} {\bibfnamefont
  {D.~R.}\ \bibnamefont {Nelson}}, \ and\ \bibinfo {author} {\bibfnamefont
  {M.~J.}\ \bibnamefont {Bowick}},\ }\href@noop {} {\bibfield  {journal}
  {\bibinfo  {journal} {Physical Review Letters}\ }\textbf {\bibinfo {volume}
  {122}},\ \bibinfo {pages} {048001} (\bibinfo {year} {2019})}\BibitemShut
  {NoStop}%
\bibitem [{\citenamefont {Sadik}\ and\ \citenamefont
  {Dias}(2021)}]{sadik2021local}%
  \BibitemOpen
  \bibfield  {author} {\bibinfo {author} {\bibfnamefont {S.}~\bibnamefont
  {Sadik}}\ and\ \bibinfo {author} {\bibfnamefont {M.~A.}\ \bibnamefont
  {Dias}},\ }\href@noop {} {\bibfield  {journal} {\bibinfo  {journal} {Journal
  of the Mechanics and Physics of Solids}\ }\textbf {\bibinfo {volume} {151}},\
  \bibinfo {pages} {104370} (\bibinfo {year} {2021})}\BibitemShut {NoStop}%
\bibitem [{\citenamefont {Choi}\ \emph {et~al.}(2019)\citenamefont {Choi},
  \citenamefont {Dudte},\ and\ \citenamefont {Mahadevan}}]{Choi2019}%
  \BibitemOpen
  \bibfield  {author} {\bibinfo {author} {\bibfnamefont {G.~P.}\ \bibnamefont
  {Choi}}, \bibinfo {author} {\bibfnamefont {L.~H.}\ \bibnamefont {Dudte}}, \
  and\ \bibinfo {author} {\bibfnamefont {L.}~\bibnamefont {Mahadevan}},\
  }\href@noop {} {\bibfield  {journal} {\bibinfo  {journal} {Nature Materials}\
  }\textbf {\bibinfo {volume} {18}},\ \bibinfo {pages} {999} (\bibinfo {year}
  {2019})}\BibitemShut {NoStop}%
\bibitem [{\citenamefont {Choi}\ \emph {et~al.}(2021)\citenamefont {Choi},
  \citenamefont {Dudte},\ and\ \citenamefont {Mahadevan}}]{Choi2020}%
  \BibitemOpen
  \bibfield  {author} {\bibinfo {author} {\bibfnamefont {G.~P.}\ \bibnamefont
  {Choi}}, \bibinfo {author} {\bibfnamefont {L.~H.}\ \bibnamefont {Dudte}}, \
  and\ \bibinfo {author} {\bibfnamefont {L.}~\bibnamefont {Mahadevan}},\
  }\href@noop {} {\bibfield  {journal} {\bibinfo  {journal} {Physical Review
  Research}\ }\textbf {\bibinfo {volume} {3}},\ \bibinfo {pages} {043030}
  (\bibinfo {year} {2021})}\BibitemShut {NoStop}%
\bibitem [{\citenamefont {Love}(2013)}]{love2013treatise}%
  \BibitemOpen
  \bibfield  {author} {\bibinfo {author} {\bibfnamefont {A.~E.~H.}\
  \bibnamefont {Love}},\ }\href@noop {} {\emph {\bibinfo {title} {A treatise on
  the mathematical theory of elasticity}}}\ (\bibinfo  {publisher} {Cambridge
  university press},\ \bibinfo {year} {2013})\BibitemShut {NoStop}%
\bibitem [{\citenamefont {Grosberg}\ and\ \citenamefont
  {Khokhlov}(1994)}]{Grosberg}%
  \BibitemOpen
  \bibfield  {author} {\bibinfo {author} {\bibfnamefont {A.~Y.}\ \bibnamefont
  {Grosberg}}\ and\ \bibinfo {author} {\bibfnamefont {A.~R.}\ \bibnamefont
  {Khokhlov}},\ }\href@noop {} {\emph {\bibinfo {title} {Statistical physics of
  macromolecules}}}\ (\bibinfo  {publisher} {Amer Inst of Physics},\ \bibinfo
  {year} {1994})\BibitemShut {NoStop}%
\bibitem [{\citenamefont {Inglis}(1913)}]{inglis1913stresses}%
  \BibitemOpen
  \bibfield  {author} {\bibinfo {author} {\bibfnamefont {C.~E.}\ \bibnamefont
  {Inglis}},\ }\href@noop {} {\bibfield  {journal} {\bibinfo  {journal} {Trans
  Inst Naval Archit}\ }\textbf {\bibinfo {volume} {55}},\ \bibinfo {pages}
  {219} (\bibinfo {year} {1913})}\BibitemShut {NoStop}%
\bibitem [{\citenamefont {Zehnder}\ and\ \citenamefont
  {Potdar}(1998)}]{zehnder1998williams}%
  \BibitemOpen
  \bibfield  {author} {\bibinfo {author} {\bibfnamefont {A.~T.}\ \bibnamefont
  {Zehnder}}\ and\ \bibinfo {author} {\bibfnamefont {Y.~K.}\ \bibnamefont
  {Potdar}},\ }\href@noop {} {\bibfield  {journal} {\bibinfo  {journal}
  {International Journal of Fracture}\ }\textbf {\bibinfo {volume} {93}},\
  \bibinfo {pages} {409} (\bibinfo {year} {1998})}\BibitemShut {NoStop}%
\bibitem [{\citenamefont {Han}\ \emph {et~al.}(2021)\citenamefont {Han},
  \citenamefont {Lewicka},\ and\ \citenamefont {Mahadevan}}]{HLM2021}%
  \BibitemOpen
  \bibfield  {author} {\bibinfo {author} {\bibfnamefont {Q.}~\bibnamefont
  {Han}}, \bibinfo {author} {\bibfnamefont {M.}~\bibnamefont {Lewicka}}, \ and\
  \bibinfo {author} {\bibfnamefont {L.}~\bibnamefont {Mahadevan}},\ }\href
  {https://arxiv.org/abs/2104.03486} {\bibfield  {journal} {\bibinfo  {journal}
  {arXiv preprint}\ } (\bibinfo {year} {2021})}\BibitemShut {NoStop}%
\bibitem [{\citenamefont {Yang}\ \emph {et~al.}(2021)\citenamefont {Yang},
  \citenamefont {Vella},\ and\ \citenamefont {Holmes}}]{yang2021grasping}%
  \BibitemOpen
  \bibfield  {author} {\bibinfo {author} {\bibfnamefont {Y.}~\bibnamefont
  {Yang}}, \bibinfo {author} {\bibfnamefont {K.}~\bibnamefont {Vella}}, \ and\
  \bibinfo {author} {\bibfnamefont {D.~P.}\ \bibnamefont {Holmes}},\
  }\href@noop {} {\bibfield  {journal} {\bibinfo  {journal} {Science Robotics}\
  }\textbf {\bibinfo {volume} {6}} (\bibinfo {year} {2021})}\BibitemShut
  {NoStop}%
\bibitem [{\citenamefont {Weischedel}\ \emph {et~al.}(2012)\citenamefont
  {Weischedel}, \citenamefont {Tuganov}, \citenamefont {Hermansson},
  \citenamefont {Linn},\ and\ \citenamefont
  {Wardetzky}}]{weischedel2012construction}%
  \BibitemOpen
  \bibfield  {author} {\bibinfo {author} {\bibfnamefont {C.}~\bibnamefont
  {Weischedel}}, \bibinfo {author} {\bibfnamefont {A.}~\bibnamefont {Tuganov}},
  \bibinfo {author} {\bibfnamefont {T.}~\bibnamefont {Hermansson}}, \bibinfo
  {author} {\bibfnamefont {J.}~\bibnamefont {Linn}}, \ and\ \bibinfo {author}
  {\bibfnamefont {M.}~\bibnamefont {Wardetzky}},\ }\href@noop {} {\  (\bibinfo
  {year} {2012})}\BibitemShut {NoStop}%
\bibitem [{\citenamefont {van Rees}\ \emph {et~al.}(2017)\citenamefont {van
  Rees}, \citenamefont {Vouga},\ and\ \citenamefont
  {Mahadevan}}]{vanrees2017growth}%
  \BibitemOpen
  \bibfield  {author} {\bibinfo {author} {\bibfnamefont {W.~M.}\ \bibnamefont
  {van Rees}}, \bibinfo {author} {\bibfnamefont {E.}~\bibnamefont {Vouga}}, \
  and\ \bibinfo {author} {\bibfnamefont {L.}~\bibnamefont {Mahadevan}},\ }\href
  {\doibase 10.1073/pnas.1709025114} {\bibfield  {journal} {\bibinfo  {journal}
  {Proceedings of the National Academy of Sciences}\ }\textbf {\bibinfo
  {volume} {114}},\ \bibinfo {pages} {11597} (\bibinfo {year} {2017})},\
  \Eprint
  {http://arxiv.org/abs/https://www.pnas.org/content/114/44/11597.full.pdf}
  {https://www.pnas.org/content/114/44/11597.full.pdf} \BibitemShut {NoStop}%
\bibitem [{\citenamefont {Peyr{\'e}}(2011)}]{peyre2011numerical}%
  \BibitemOpen
  \bibfield  {author} {\bibinfo {author} {\bibfnamefont {G.}~\bibnamefont
  {Peyr{\'e}}},\ }\href@noop {} {\bibfield  {journal} {\bibinfo  {journal}
  {IEEE Computing in Science and Engineering}\ }\textbf {\bibinfo {volume}
  {13}},\ \bibinfo {pages} {94} (\bibinfo {year} {2011})}\BibitemShut {NoStop}%
\bibitem [{\citenamefont {Kimmel}\ and\ \citenamefont
  {Sethian}(1998)}]{kimmel1998computing}%
  \BibitemOpen
  \bibfield  {author} {\bibinfo {author} {\bibfnamefont {R.}~\bibnamefont
  {Kimmel}}\ and\ \bibinfo {author} {\bibfnamefont {J.~A.}\ \bibnamefont
  {Sethian}},\ }\href@noop {} {\bibfield  {journal} {\bibinfo  {journal}
  {Proceedings of the National Academy of Sciences}\ }\textbf {\bibinfo
  {volume} {95}},\ \bibinfo {pages} {8431} (\bibinfo {year}
  {1998})}\BibitemShut {NoStop}%
\end{thebibliography}%
\end{document}